# Near-infrared light-induced superconducting-like state in underdoped YBa$_2$Cu$_3$O$_y$ studied by *c*-axis terahertz third-harmonic generation

Kota Katsumi[1,2], Morihiko Nishida[1], Stefan Kaiser[3,4], Shigeki Miyasaka[5], Setsuko Tajima[5], and Ryo Shimano[1,2]

[1] *Department of Physics, The University of Tokyo, Hongo, Tokyo, 113-0033, Japan*

[2] *Cryogenic Research Center, The University of Tokyo, Yayoi, Tokyo, 113-0032, Japan*

[3] *Max Planck Institute for Solid State Research, Stuttgart, 70569, Germany*

[4] *Institute of Solid State and Materials Physics, Technical University Dresden, Dresden, 01062, Germany*

[5] *Department of Physics, Osaka University, Toyonaka, Osaka, 560-0043, Japan*

**Abstract**

Recent observation of the light-induced superconducting (SC)-like transient response in the *c*-axis optical conductivity far above the SC transition temperature $T_c$ in underdoped YBa$_2$Cu$_3$O$_y$ (YBCO) has attracted great attention in the field of high-$T_c$ superconductors. Since then, various theoretical and experimental studies have been devoted to elucidating its microscopic origin. One prominent fingerprint of the light-induced superconductivity is the emergence of $1/\omega$-like spectral behavior in the imaginary part of the optical conductivity in the terahertz (THz) frequency range. However, the spectral profile can also be described by the Drude response of the quasiparticles (QPs) with a substantially low scattering rate. To circumvent this critical ambiguity, we investigated the light-induced nonequilibrium state in an underdoped YBCO sample with $T_c$ of 61 K using the nonlinear THz optical response originating from the SC collective excitation of the ac-driven Josephson current. Upon the near-infrared (NIR) photoexcitation above $T_c$ in the YBCO sample, the $1/\omega$-like spectral behavior in the imaginary part of the optical conductivity emerges, consistent with the previous studies. However, the THz third-harmonic generation arising from the ac-driven Josephson current along the *c*-axis was absent in the NIR photoexcited state. These results indicate that the NIR-pump induced state exhibiting the $1/\omega$-like response above $T_c$ is distinct from the long-range ordered SC state in equilibrium. Based on these observations, the possible origins of the irregularly coherent charge carrier response along the *c*-axis induced by the photoexcitation are discussed.

# I. INTRODUCTION

Recent advancements in ultrafast light sources and spectroscopic techniques have accelerated the study of nonequilibrium dynamics in versatile quantum materials. One of the most intriguing examples is the light-induced superconductivity in high-temperature cuprate superconductors. Upon photoexcitation, a plasma edge-like structure was observed in transient *c*-axis terahertz (THz) reflectivity spectra above the superconducting (SC) transition temperature $T_c$ in stripe-ordered La-based cuprate superconductors $La_{1.675}Eu_{0.2}Sr_{0.125}CuO_4$ [1] and $La_{2-x}Ba_xCuO_4$ [2] and interpreted as the Josephson plasma edge of the light-induced SC state. Remarkably, the light-induced SC-like behavior was also identified in underdoped $YBa_2Cu_3O_y$ (YBCO) samples that do not exhibit the stripe order, even at room temperature depending on the doping [3,4]. The light-induced SC-like state in YBCO was first reported by using mid-infrared (MIR) photoexcitation which was tuned to the resonance of apical-oxygen phonon. In the subsequent experiments, the SC-like response was also observed using near-infrared (NIR) photoexcitation [5,6], where the photon energy is far different from that of the MIR pumping case. Various theoretical [7-20] and experimental studies [21-34] have been devoted to elucidating the microscopic origin of the observed light-induced SC-like behavior. In these experiments, the transient emergence of the $1/\omega$-like spectral profile in the imaginary part of the optical conductivity $\sigma_2(\omega)$, which is identical to the response of the SC condensate in equilibrium, is interpreted as a fingerprint of the light-induced superconductivity[1-4]. However, ambiguity remains that one cannot distinguish the $1/\omega$-like response in $\sigma_2(\omega)$ from a Drude response of the quasiparticles (QPs) with a substantially low scattering rate in the limited spectral window of the THz spectroscopy [6]. Accordingly, another ultrafast probe of the SC order parameter is desired to have a deeper insight into the observed SC-like behavior.

For this purpose, we shed light on the THz nonlinear optical responses originating from the collective excitations of the SC order parameter. By virtue of the recent developments in intense THz light sources [35,36], various THz nonlinear responses associated with the SC collective modes have been investigated [37-54]. Since those collective excitations directly manifest the growth of the SC order parameter, their nonlinear responses in the THz frequency range can keep track of the dynamics of the order parameter on the ultrafast time scale.

In this study, we employed the THz third-harmonic generation (THG) arising from the ac Josephson current to probe the SC order parameter in the photoexcited nonequilibrium state of underdoped YBCO. In particular, we focus on the case of NIR photoexcitation to examine the properties of the light-induced state showing the $1/\omega$-like spectral behavior above $T_c$, and we do not argue about the MIR phonon-pumping case.

This article is organized as follows. In the Experimental results section A, we describe the results of optical pump-THz probe (OPTP) measurements on an underdoped YBCO sample ($T_c$ = 61 K) using NIR pump at 800 nm. We identified the pump-induced $1/\omega$-like spectral behavior in the imaginary part of the optical conductivity above $T_c$, which agrees with the previous reports [3,4]. In Experimental Results section B, we examined the photoexcited state using THz-THG along the $c$-axis, which acts as an indicator of the $c$-axis SC coherence [53]. While the THz-THG signal was clearly observed below $T_c$ without NIR photoexcitation, it was not detected in the photoexcited state above $T_c$, which is against the expectation for the light-induced long-range SC coherence inferred from the appearance of the $1/\omega$-like spectral behavior in the imaginary part of the optical conductivity. This result indicates that the NIR-pump induced state exhibiting the $1/\omega$-like spectral behavior is distinct from the long-range ordered SC state in equilibrium. We discuss the possible origins of the NIR pump-induced nonequilibrium state above $T_c$ in the Discussion section.

## II. EXPERIMENTAL RESULTS

### A. Photoexcited nonequilibrium state studied by OPTP spectroscopy

*1. THz transient responses in the photoexcited state below $T_c$*

A schematic of the OPTP measurement is depicted in Fig. 1(a). The details of the sample and experiments are described in Appendix A, and the equilibrium optical properties of the sample are summarized in Appendix B. Figure 1(b) shows the transient THz reflectivity along the $c$-axis of the YBCO at 20 K measured at the pump-probe delay time $t_{pp}$ = 3.2 ps with the 800-nm pump fluence of 3.4 mJ/cm$^2$. A redshift of the Josephson plasma edge is observed after the photoexcitation, indicating that the superconductivity is destroyed by the photoexcitation. The transient reflectivity change at the selected pump-probe delay times $t_{pp}$ is displayed in Fig. 1(c). For all the positive delay times, the negative reflectivity change is discerned below the equilibrium Josephson plasma resonance (JPR) frequency of 1.2 THz, which indicates the redshift of the Josephson plasma edge. It is counter-intuitive that the JPR remains, although redshifted, even after the intense photoexcitation with the fluence of 3.4 mJ/cm$^2$, which is much larger than the threshold fluence of $F_{th}$ = 10-200 μJ/cm$^2$ for the total destruction of the superconductivity in cuprate superconductors as reported in the time-resolved optical spectroscopy [55-59] and angle-resolved photoemission spectroscopy (ARPES) measurements [60-62]. This non-vanishing JPR even under the intense photoexcitation far above the threshold fluence ($F_{th}$) is because the penetration depth of the 800-nm pump pulse $d_{pump}$ = 0.22 μm [5] is much shorter than that of the THz probe pulse $d_{THz}$ = 8 μm at 1 THz estimated from the refractive index at 20 K obtained by THz time-domain spectroscopy (THz-TDS). In Appendix C, we show that when the system reaches a quasi-thermal equilibrium state, the transient optical response is reasonably reproduced by considering the temperature increase due to the pump-induced heating (heating model). Furthermore, we present in Appendix D that the transient optical response right after photoexcitation at $t_{pp}$ = 3.2 ps is consistently described by a model assuming that the superconductivity is destroyed in the photoexcited surface region deeper than the pump penetration depth, whose optical constants are described by that at 60 K (destroyed layer model).

## *2. THz transient responses in the photoexcited state above $T_c$*

Now we examine the photoexcited nonequilibrium state of the YBCO above $T_c$. Considering the penetration depth mismatch between the pump and probe pulses, we extracted the transient optical response at the surface region of the sample (the details of the analysis are described in Appendix E). In Fig. 2, we plot the real and imaginary parts of the optical conductivity at the photoexcited surface region at 100 K. As shown in Fig. 2(b), the imaginary part of the optical conductivity $\sigma_2(\omega)$ measured at $t_{pp}$ = 0.8 ps (red circles) exhibits the $1/\omega$-like behavior, which coincides well with that at 40 K in equilibrium without the photoexcitation (blue dashed curve). Concomitantly, the real part of the optical conductivity $\sigma_1(\omega)$ exhibits a slight increase as plotted in Fig. 2(a). Although the spectral feature of $\sigma_2(\omega)$ at $t_{pp}$ = 0.8 ps is almost identical to that in SC state, the pump-induced increase in $\sigma_1(\omega)$ and $\sigma_2(\omega)$ at $t_{pp}$ = 0.8 ps can be simultaneously fitted by the Drude model:

$$\Delta\sigma(\omega) = \frac{i\varepsilon_0 \omega_p^2}{\omega + i\gamma}. \tag{1}$$

Here, $\varepsilon_0$ is the vacuum permittivity, and $\omega_p$ and $\gamma$ are the plasma angular frequency and scattering rate, respectively. Figures 2(a) and (b) also show the fitting curves to the optical conductivity at $t_{pp}$ = 0.8 ps with $\omega_p/2\pi$ = 4.7 THz and $\gamma/2\pi$ = 0.14 THz. Note that it has recently been pointed out that the saturation effect on the pump-induced refractive index change along the depth direction should be taken into account adequately to extract the true surface optical spectra [63]. In the present case, however, the squared plasma frequency $\omega_p^2$ evaluated by the Drude model follows the pump fluence and does not show a saturation behavior, indicating that such effect is not substantial. Similar to the result at $t_{pp}$ = 0.8 ps, the pump-induced change in $\sigma_1(\omega)$ and $\sigma_2(\omega)$ measured at $t_{pp}$ = 1.3 ps can be simultaneously fitted by the Drude model with $\omega_p/2\pi$ = 6.9 THz and $\gamma/2\pi$ = 1.0 THz, as shown by the green circles in Figs. 2(c) and (d). Figure 2(e) summarizes the time evolution of the pump-induced change in the imaginary part of the optical conductivity $\Delta\sigma_2(\omega)$. The vertical arrows denote the peaks in $\Delta\sigma_2(\omega)$ spectra. The peak frequency increases with the delay time $t_{pp}$, which is also identified in the previous study [26]. However, it is challenging to distinguish whether the observed transient increase in $\sigma_2(\omega)$ reflects the SC or Drude response due to the limited frequency range of the THz spectroscopy. To clarify this problem, we examine the photoexcited nonequilibrium state in YBCO using the THz-THG arising from the ac-Josephson current along the *c*-axis.

Before moving to the THz-THG experiments, we further investigate the temperature dependence of the light-induced transient THz response. To study the temperature dependence of the light-induced plasma-edge-like response above $T_c$ with a better signal-to-noise ratio, we focus on the pump-induced change in the THz probe electric field (*E*-field) instead of the transient optical conductivity spectrum as follows. Figure 3(a) shows the 800-nm pump-induced change in the THz *E*-field $\Delta E(t_{gate})$ above $T_c$ as a function of the probe-gate delay time ($t_{gate}$) with the pump-probe delay time fixed at $t_{pp} = 0.8$ ps. The corresponding power spectra of $\Delta E(t_{gate})$, denoted by $\Delta I$, are plotted in Fig. 3(b). The spectral weight of $\Delta I$ integrated from 0.4 THz to 1.0 THz divided by the reflected THz power without photoexcitation, $\int d\omega \frac{\Delta I}{I}$, is displayed in Fig. 3(c) as a function of the pump-probe delay time $t_{pp}$ (magenta, left axis). The dynamics of frequency-integrated $\Delta I/I$ follows the time evolution of $[\omega\sigma_2(\omega)]^2$ at $\omega/2\pi = 0.4$ THz obtained using the single layer model analysis described in Appendix E (light blue, right axis). Thus, the frequency-integrated intensity $\Delta I/I$ can be used as an indicator of the pump-induced increase in $\sigma_2(\omega)$. We present the temperature dependence of $\Delta I/I$ in Fig. 3(d). Here, the black horizontal line denotes the noise floor of $\Delta I/I$ at 300 K estimated from the data at the negative delay, $t_{pp} = -9.6$ ps. As the temperature increases, the integrated intensity $\Delta I/I$ decreases and approaches the noise floor around 210 K which is consistent with the previous study of OPTP measurements for similarly doped YBCO [26]. Notably, this temperature is close to the pseudogap opening temperature $T^*$ of similarly doped YBCO, reported as 235±10 K in neutron scattering [64], while lower than 260±40 K in Nernst measurement [65] and 270±20 K in optical spectroscopy measurement [66]. We will discuss this point later in the Discussion section.

In the following, to gain a deeper insight into the origin of the NIR pump-induced $1/\omega$-like behavior in $\sigma_2(\omega)$, we investigate the THz-THG arising from the SC collective excitation in the photoexcited state.

### B. Photoexcited nonequilibrium state studied by THz THG

#### 1. THz-THG measurements

In Fig. 4(a), we present the power spectra of the reflected narrowband THz *E*-field polarized along the *c*-axis of the YBCO at the selected temperatures. The third harmonic (TH) component is clearly identified at 1.5 THz below $T_c$. Figures 4(b) and (c) show the integrated fundamental harmonic (FH) and TH intensities as a function of the incident THz peak *E*-field ($E_{in}$) at 20 K and 60 K, respectively. Here, we integrate the FH intensity from 0.3 to 0.7 THz and the TH intensity from 1.2 to 1.8 THz. At 20 K, the TH intensity follows $E_{in}^6$ and the FH intensity follows $E_{in}^2$, indicating that the observed TH component originates from the THG process in the SC state. In contrast, at 60 K, the integrated intensity around 1.5 THz (denoted as TH in Fig. 4(c) for convenience) follows $E_{in}^2$. This 1.5 THz component at 60 K is attributed to the finite leakage of the FH component, which transmits the 1.5-THz bandpass filters (BPFs) placed after the sample. We note that the integrated TH intensity for the lowest $E$-field deviates from the expected power law dependence displayed by solid lines in both Figs. 4(b) and (c), because the signal level is close to the noise floor, which is estimated by integrating the power spectra from 2.5 to 3.1 THz, i.e., outside the TH frequency range [67,68].

To discuss the temperature dependence of the TH intensity, we evaluate the third-order nonlinear susceptibility normalized by the FH penetration depth $\chi^{(3)}_{dep}$ [69,70]. Here, we subtract the leakage of the FH intensity from the TH intensity and consider the Fresnel reflection loss of the incident FH $E$-field at the sample surface. The details of the analysis for $\chi^{(3)}_{dep}$ are described in Appendix G. The temperature dependence of $\chi^{(3)}_{dep}$ is plotted by the red circles in Fig. 4(d), showing a good agreement with that of the superfluid density $n_s$ represented by the gray circles, which we obtained by applying the two-fluid model to the complex optical conductivity in equilibrium (see Appendix B for details). This agreement is consistent with the temperature dependence expected for the THG mediated by the THz-driven ac-Josephson current, which is proportional to the superfluid density [53]. Therefore, the observed THG is attributed to the ac-Josephson current driven by the intense THz pulse polarized along the $c$-axis and serves as the ultrafast probe of the SC order parameter under the photoexcitation. In the following, we utilize this THz-THG signal to investigate the light-induced nonequilibrium state.

## 2. OP-THz THG measurements

Figure 5(a) illustrates the schematic of the OP-THG measurements for the YBCO in reflection geometry. In Fig. 5(b), we show the power spectra of the reflected narrowband THz pulse at 20 K when the 800-nm photoexcitation is "On" (blue) and "Off" (gray) with the pump-probe delay time of $t_{pp}$ = 3.2 ps. The frequency-integrated TH intensity $I_{TH}$ (integrated from 1.2 to 1.8 THz) decreases by 5.9% when the photoexcitation is "On" compared to that when the photoexcitation is "Off." Following the destroyed layer model described in Appendix D, the superconductivity is destroyed within the photoexcited surface depth of 0.94 μm with the pump fluence of 3.4 mJ/cm². By calculating the FH THz $E$-field penetrating into the SC bulk in the photoexcited state and considering the Fresnel transmission coefficients for the TH component from the SC bulk to the photoexcited surface layer and that from the surface layer to the air, the TH intensity is estimated to decrease by 5.3% compared to that without photoexcitation, showing a reasonable agreement with the experimental result. Figure 5(c) plots the pump-induced change in the TH intensity $\Delta I_{TH}$ at 20 K at the selected delay times. For all the delay times, the TH intensity is reduced, and no enhancement is identified within the error bars. These results reinforce that the NIR photoexcitation destroys the superconductivity below $T_c$.

Next, we examine the light-induced nonequilibrium state above $T_c$ by the OP-THG measurements. As described in the Experimental results section A, in the OPTP measurements at 100 K with the pump fluence of 3.4 mJ/cm$^2$, we observed the $1/\omega$-like spectral profile in $\sigma_2(\omega)$ at $t_{pp}$ = 0.8 ps, which is considerably identical to that in the equilibrium SC state at 40 K. If this transient state can be recognized as the 40 K-superconductivity, one should expect a finite THG at 100 K after photoexcitation.

First, to estimate the expected TH intensity in the photoexcited nonequilibrium state at 100 K, we consider the penetration depth mismatch between the 800-nm pump and 0.5-THz pulses, schematically drawn in Fig. 6(a). Assuming that the photoexcited surface region of the sample turns to the SC state, whose superfluid density $n_s$ is the same as that at 40 K, that surface region should have the same third-order nonlinear susceptibility $\chi^{(3)}_{dep}$ as that at 40 K. Since the 800-nm pump beam diameter is 2 mm and sufficiently larger than that of the THz pulse (0.55 mm), the volume of the SC region should be simply proportional to the pump penetration depth $d_{pump}$. Here, the penetration depth at 800 nm is set to $d_{pump}$ = 0.22 μm from the literature [5]. We calculate the penetration depth for 0.5 THz at 40 K as $d_{THz}$ = 5.7 μm using the refractive index obtained in the THz-TDS measurements. Hence, the ratio of the SC volume at 100 K in the photoexcited state measured at $t_{pp}$ = 0.8 ps to that at 40 K in equilibrium is estimated as $d_{pump}/d_{THz}$ = 3.9%.

Second, we evaluate the FH $E$-field penetrating inside the sample, as discussed in Appendix G. For the FH $E$-field inside the sample at 100 K after photoexcitation, we consider the effect of the multiple reflections inside the photoexcited surface region. Using the pump-induced surface refractive index at 100 K measured at $t_{pp}$ = 0.8 ps and refractive index at 40 K in equilibrium, the ratio of $B_{THG}(\omega)$ (defined by Eq. (G5) in Appendix G) at 0.5 THz is computed as 44. Combining this result with the SC volume ratio, we estimate that $(d_{pump}/d_{THz})^2 B_{THG}(\omega)$ = $(0.039)^2 \times 44$ = 6.7% of the TH intensity at 40 K without the photoexcitation should be observed at 100 K in the photoexcited state at $t_{pp}$ = 0.8 ps.

In the upper panel of Fig. 6(b), we present the 800-nm pump-induced change in the TH power spectrum $\Delta I_{TH}$ at 100 K measured at $t_{pp}$ = 0.8 ps and the TH power spectrum $I_{TH}$ at 40 K without the photoexcitation multiplied by 6.7%. Importantly, the pump-induced change in the TH signal, $\Delta I_{TH}$, is much smaller than the expected TH intensity despite the emergence of $1/\omega$-like spectral profile in $\sigma_2(\omega)$ upon the photoexcitation. While a slight increase in the TH intensity within the error bars is discerned at 1.33 THz (see the lower panel of Fig. 6(b) for the enlarged view), similar or even larger increases are identified at different delay times, and the center frequency completely deviates from the TH peak below $T_c$, and thus we conclude that it is within the range of noise floor.

Figure 6(c) displays the pump-induced change in the FH power spectrum ($\Delta I_{FH}$, upper panel) and the TH power spectrum ($\Delta I_{TH}$, lower panel) as a function of the delay time $t_{pp}$ measured at 100 K. Although $\Delta I_{FH}$ increases after the photoexcitation, $\Delta I_{TH}$ does not exhibit a measurable change at any delay time. Furthermore, we simulate the time evolution of the THz TH intensity after the photoexcitation by assuming that the SC order is transiently induced by the photoexcitation, as shown in Fig. 6(d) (see Appendix H for the simulation details). The simulated THz TH intensity follows the dynamics of the pump-induced SC order represented in the inset of Fig. 6(d), whereas no such signal was observed in the experiments. These results contradict the interpretation that the $1/\omega$-like spectral profile in the transient $\sigma_2(\omega)$ spectrum is attributed to the superconductivity.

### III. DISCUSSION

In our study, we do not identify the THz-THG signal from the Josephson current along the *c*-axis in the NIR photoexcited nonequilibrium state above $T_c$, while it was observed in the equilibrium SC phase below $T_c$. Hence, it is hard to attribute the NIR pump-induced $1/\omega$-like spectral response observed in the imaginary part of the optical conductivity $\sigma_2(\omega)$ to the SC phase in equilibrium. Here, we should note that the $1/\omega$-like response in $\sigma_2(\omega)$ previously reported under *the MIR pump excitation* in YBCO may have a different origin for the following reasons. Firstly, the MIR pump-induced $1/\omega$-like response shows a clear resonance when the pump photon energy is tuned to the apical-oxygen phonon energy [3-5]. Secondly, the pump-induced change in the real part of the optical conductivity $\sigma_1(\omega)$ with MIR pump is smaller than that induced by NIR pump [5]. Recently, in the case of the MIR pump excitation, the parametric amplification of the Josephson plasmons mediated by the resonantly driven phonon at 80 meV has been proposed as a mechanism for the light-induced superconductivity [17,20,34]. This parametric process might be one possible scenario to explain the $1/\omega$-like transient response in $\sigma_2(\omega)$ in the present study with 800 nm (1.55 eV) NIR optical pump if the NIR pump pulses can drive the phonon at 80 meV off-resonantly. As presented in Appendix F, however, the $1/\omega$-like response in $\sigma_2(\omega)$ is also observed when the 800-nm pump is polarized along the *a*-axis, making a stark contrast to the previous work with the MIR pump where the $1/\omega$-like response was observed only for the *c*-axis pump [26]. This polarization dependence indicates that the apical oxygen phonon is irrelevant in the case of the NIR photoexcitation. Furthermore, the spectral width of our Gaussian pump pulse (800 nm, 100 fs) is 4.4 THz (18 meV), which is too small to excite the 80-meV phonon through the impulsive-stimulated Raman process. It should be also noted that in the proposed framework of the parametric process, the superconductivity should be transiently enhanced even below $T_c$ [17,20,34]. By contrast, our experimental results demonstrate that the 800-nm photoexcitation destroys the superconductivity along the *c*-axis below $T_c$. Therefore, the present result of NIR photoexcitation is unlikely attributed to the parametrically amplified plasmon process.

It is highly nontrivial that the photoexcited state in underdoped YBCO far above $T_c$ exhibits such a coherent carrier transport along the *c*-axis with a substantially low scattering rate, less than 0.15 THz, albeit with the incoherent *c*-axis transport in equilibrium [71]. Even though the light-induced state lacks the SC collective modes, it is still tempting to infer that the observed coherent charge carrier response is related to superconductivity. Following this scenario, the phase-fluctuating SC state without the long-range order may be a candidate to explain the lack of the THz-THG associated with superconductivity. In this view, it is indicative that the coherent Drude responses have been observed previously in equilibrium along the *c*-axis far above $T_c$ [72], and for the in-plane response even below $T_c$ [73], both of which were interpreted as a signature of the phase-fluctuating superconductivity. It is also notable that a recent theoretical study has shown that photoexcitation enhances the *d*-wave SC correlation but with a substantially short pairing correlation length [74]. Though it assumes the in-plane photoexcitation of the stripe-ordered cuprates, the enhanced superconductivity with substantially short-range correlation may be relevant to our scenario of the light-induced phase-fluctuating superconductivity.

Another possible scenario is that the observed SC-like behavior in $\sigma_2(\omega)$ is associated with the pseudogap because the 800-nm pump-induced THz reflectivity change persists up to the pseudogap opening temperature $T^*$, as mentioned in the Results section A-2. While we cannot exclude the possibility of a simple coincidence, there are two reasons why we presume that this onset temperature is possibly related to the pseudogap. First, the light-induced transient increase in $\sigma_2(\omega)$ has also been reported in variously underdoped YBCO samples up to $T^*$ with the MIR [3,4,26] and NIR pulse excitation [6]. Second, previous optical spectroscopy [66,75-78] and theoretical studies [79-81] have revealed that the equilibrium *c*-axis optical response of cuprates is sensitive to the electronic density of states in the antinodal region where the pseudogap opens: the interlayer hopping integral for cuprates strongly depends on the *ab*-plane momentum **k** of carriers as $t_\perp(\mathbf{k}) \propto [\cos(k_x a_0) - \cos(k_y b_0)]^2$ ($a_0$ and $b_0$ are the lattice constants for the *a* and *b* axis, respectively) [79,80], which is maximum at the antinode. Accordingly, the light-induced SC-like behavior in $\sigma_2(\omega)$ might be related to the pseudogap in the present case of YBCO.

Though the origin of the pseudogap is still under intensive debate [78,82-85], some theoretical studies have recently suggested the presence of a pair-density wave (PDW) state at high temperature above $T_c$ and its relation with the pseudogap has been discussed [86-88]. The existence of the PDW has been reported by the scanning tunneling microscopy (STM) measurements in Bi2212 far below $T_c$ [89,90] and even in the high-temperature regime up to $1.5T_c$ [91]. Interestingly, it has been theoretically proposed that the photoexcited state of the PDW can exhibit the $1/\omega$-like response in $\sigma_2(\omega)$ without showing the Meissner effect [18]. This scenario seems to be consistent with our experimental results that the THz-THG mediated by the ac-Josephson current is absent in the photoexcited state out of equilibrium.

## IV. CONCLUSION

We studied the NIR pump-induced nonequilibrium state in the underdoped YBCO sample. First, we performed the OPTP spectroscopy for the YBCO along the *c*-axis. We observed the redshift of the Josephson plasma edge in the transient THz reflectivity spectrum after the 800-nm photoexcitation below $T_c$, indicating the light-induced destruction of superconductivity. On the other hand, we identified the $1/\omega$-like response in the imaginary part of the optical conductivity $\sigma_2(\omega)$ immediately after the photoexcitation above $T_c$ up to the pseudogap temperature, consistent with the previous reports [3,4,6,26].

Next, we investigated the NIR photoexcited nonequilibrium state using the THz-THG signal arising from ac-driven Josephson current along the *c*-axis, which acts as an indicator of the SC order parameter. Below $T_c$, the THz-THG intensity displayed a reduction upon the photoexcitation, confirming that the photoexcitation destroys the superconductivity. On the other hand, the THz-THG signal from the ac-Josephson current was not identified in the NIR photoexcited nonequilibrium state of the YBCO above $T_c$, despite the appearance of the $1/\omega$-like transient response in the imaginary part of the *c*-axis optical conductivity. This result indicates that the light-induced state that exhibits the $1/\omega$-like response along the *c*-axis imaginary part of the optical conductivity is distinct from the equilibrium SC state below $T_c$. This result leads to a question on the microscopic origin of the unusually coherent charge carrier response along the *c*-axis that appeared in the photoexcited state of underdoped YBCO above $T_c$, which deserves further experimental and theoretical investigation and will provide important insights for the understanding of electronic states above $T_c$ in underdoped cuprate superconductors. It will be also promising to apply the study of nonlinear THz responses arising from the collective modes to the case of the *MIR-phonon pumped* nonequilibrium state, where the optical responses are reported to be qualitatively different from the case of the NIR pumping.

## ACKNOWLEDGEMENTS

We acknowledge N. Tsuji and Y. Gallais for fruitful discussions. This work was partly supported by JSPS KAKENHI Grants No. 18H05324 and No. 15H02102 and JST CREST Grant No. JPMJCR19T3. K.K. was supported by JSPS Research Fellowship for Young Scientists (Grant No. 19J12873). S.K. acknowledges funding by the European Union (ERC, T-Higgs, GA 101044657).

We also acknowledge the Max Planck-UBC-U Tokyo Center for Quantum Materials for the collaboration.

## APPENDIX A: DETAIS OF SAMPLE AND EXPERIMENTS

First, we performed optical pump and THz probe (OPTP) measurements using an underdoped YBCO single crystal ($T_c$ = 61 K) grown by the pulling method [92]. The SC transition temperature was determined by the magnetic susceptibility measurement using a SC quantum interference device (SQUID). Figure 7 shows the magnetic moments for the YBCO single crystal measured by SQUID under zero-field cooling. The SC transition temperature $T_c$ is determined by the onset of the drop in the magnetic moment. The crystal was cut and polished to give an *ac* surface whose size is $a \times b \times c$ = 3 × 1 × 2 mm$^3$.

Figure 1(a) illustrates a schematic of the OPTP spectroscopy. The output of a regenerative amplified Ti:sapphire laser system with the central wavelength of 800 nm, pulse duration of 100 fs, pulse energy of 4 mJ, and repetition rate of 1 kHz, was divided into three beams: one for the generation of the THz probe pulse, one for the optical pump pulse, and the other for the gate pulse for the electro-optic (EO) sampling. The THz probe pulse was generated by the optical rectification in a ZnTe (110) crystal. The reflected THz pulse from the sample was detected by the EO sampling in a 2-mm-thick ZnTe (110) crystal. We present the OPTP results in the Experimental Results section A.

Next, optical pump-THG probe (OP-THG) measurements were performed in reflection geometry for the same YBCO sample. We generated an intense narrowband THz pulse employing the tilted-pulse-front technique with a LiNbO$_3$ crystal [35,36]. To obtain the narrowband THz pulse with the center frequency of 0.5 THz, we inserted BPFs in the incident THz beam pass. The peak *E*-field of the narrowband THz pulse was estimated as 25 kV/cm by EO sampling in the 380-μm-thick GaP (110) crystal placed inside the cryostat. To extract the TH components of the reflected THz pulse, 1.5 THz BPFs were inserted in the reflected THz beam pass so that the FH component at 0.5 THz was eliminated. The reflected THz pulse from the sample was detected by EO sampling in the 2-mm-thick ZnTe (110) crystal. The OP-THG results are displayed in the Experimental Results section B.

## APPENDIX B: EQUILIBRIUM OPTICAL CONSTANTS

In Figs. 8(a)-(d), we present the equilibrium optical constants of the YBCO sample along *c*-axis measured by THz-TDS in reflection geometry. As shown in Fig. 8(a), the reflectivity spectrum displays the plasma edge associated with the JPR below $T_c$ = 61 K. As the temperature is lowered, the Josephson plasma edge shows a blueshift and approaches 1.2 THz at 20 K, manifesting the growth of the *c*-axis SC coherence. Concomitantly, a peak appears in the loss function spectrum −Im(1/$\varepsilon(\omega)$) at the JPR frequency as plotted in Fig. 8(b). Figures 8(c) and (d) show that below $T_c$, the spectral weight of $\sigma_1(\omega)$ is suppressed, and $\sigma_2(\omega)$ exhibits an increasing tendency toward lower frequency described by the 1/$\omega$-profile, reflecting the development of the SC phase stiffness. These spectral features agree with the previous studies on the *c*-axis optical response of YBCO [76,93].

Since the frequency range of the obtained optical constants is limited above 0.7 THz due to the sample size, the optical constants below 0.7 THz are estimated by fitting the complex optical conductivity $\sigma(\omega) = -i\varepsilon_0\omega(\varepsilon(\omega)-\varepsilon_b)$ with the following two-fluid model [5,6,94]:

$$\varepsilon(\omega) = \varepsilon_b - \frac{n_s}{\omega^2} - \frac{\omega_d^2}{\omega^2 + i\gamma_d\omega} - \sum_j \frac{S_j^2}{\omega^2 - \omega_j^2 + i\gamma_j\omega}. \tag{B1}$$

Here, $n_s$ is superfluid density and set to zero above $T_c$. $\omega_d$ and $\gamma_d$ are the plasma angular frequency and the scattering rate of the Drude component, respectively. The scattering rate $\gamma_d$ is fixed to 10 THz, which is a typical value for the similarly doped YBCO [6,94]. The parameters of $\omega_j$, $\gamma_j$, and $S_j$ are the central angular frequency, scattering rate, and oscillator strength of the *j*-th oscillator representing the infrared active phonons, respectively. We adopt the parameters of these oscillators taken from Ref. [94] and fixed at all temperatures. $\varepsilon_b$ is the background dielectric constant of YBCO and set to 4.5 from the literature value [95].

The fitting results for the real and imaginary parts of the optical conductivity are shown in Figs. 8(c) and (d), respectively. Whereas the fitting reasonably reproduces the imaginary part of the optical conductivity $\sigma_2(\omega)$, the real part $\sigma_1(\omega)$ slightly deviates from the fitting, possibly because the phonon parameters are fixed. Figures 8(e) and (f) plots the temperature dependence of the fitting parameters $n_s$ and $\omega_d^2$, respectively. The superfluid density $n_s$ displays an increase below $T_c$, manifesting the growth of the SC phase stiffness.

## APPENDIX C: THE HEATING MODEL SIMULATION

To extract the transient *c*-axis THz response at the photoexcited surface region of YBCO samples, various models have been utilized which suppose that the light-induced refractive index change relaxes along the sample depth direction within the pump penetration depth [3-6,26]. However, it was pointed out that such models can give rise to artifacts when one extracts the optical constants of the photoexcited surface region, particularly in the SC state [32]. In fact, it has been demonstrated that, below $T_c$, the transient optical constants in La$_{2-x}$Sr$_x$CuO$_4$ are well described by the pump-induced sample heating caused by the injected pump energy in particular for pump-probe delay times when the system reaches a quasi-thermal equilibrium state [32]. Since the typical electron-phonon scattering time in cuprate is ~ 6 ps [96], it is reasonable to apply this heating model to the present case of YBCO after $t_{pp}$ = 6 ps.

Accordingly, we analyze the transient optical constants at $t_{pp}$ = 12.8 ps using this heating model reported in the previous work of Ref. [32] as follows. Assuming that the optical pump fluence $F_0$ decays exponentially inside the sample with the penetration depth $d_{pump}$ as $F(z) = F_0\exp(-z/d_{pump})$, one can calculate the pump energy density absorbed at the depth $z$ from the surface as

$$I(z) = (1-R)\left(-\frac{dF(z)}{dz}\right) = (1-R)\frac{F_0}{d_{pump}}\exp\left(-\frac{z}{d_{pump}}\right), \tag{C1}$$

where $F_0$ is the pump fluence, and $R$ is the reflectivity at the pump wavelength. In the case of 800-nm pump pulse, $R$ = 0.1 and $d_{pump}$ = 0.22 µm from the literature [5]. If the energy of $I(z)$ is absorbed by the sample, the temperature at the depth $z$ of the sample $T_f(z)$ can be connected to $I(z)$ by the following integral equation:

$$I(z) = \int_{T_i}^{T_f(z)} dT N C_s(T). \tag{C2}$$

Here, $T_i$ is the initial temperature of the sample, $C_s(T)$ is the specific heat of the sample, and $N$ is the number of molecules in the excited volume. In the case of YBCO, it is known that the specific heat $C_s(T)$ can be described by the following equation over a wide temperature range [5,97,98]:

$$C_s(T) = \gamma_s T + \beta_s T^3, \tag{C3}$$

where $\gamma_s$ = 2.3 mJ mol$^{-1}$ K$^{-2}$ and $\beta_s$ = 0.394 mJ mol$^{-1}$ K$^{-4}$ are the electronic and lattice coefficients to the specific heat taken from the literature values [98]. We calculate $T_f(z)$ for a given $F_0$ by numerically solving Eq. (C2) with Eqs. (C1) and (C3). Figure 9(a) shows the spatial distribution of the pump fluence $F(z)$ (magenta, left axis) and the sample temperature in a quasi-equilibrium $T_f(z)$ (light blue, right axis) obtained by the heating model simulation for the incident pump fluence of 3.4 mJ/cm². One can see that the temperature rise penetrates deeper into the sample than the pump penetration depth.

Using the temperature dependence of the complex refractive index (we present the real part $n_1(\omega)$ in Fig. 9(b) and the imaginary part $n_2(\omega)$ in Fig. 9(c) for selected frequencies), the computed $T_f(z)$ is converted to the spatial distribution of the complex refractive index as shown in Figs. 9(d) and (e). Lastly, using the obtained spatial distribution of the complex refractive index, we calculate the effective optical response by dividing the sample surface region into layers [3,4,32]. In this study, we divided the sample surface of thickness 30 µm into 3000 layers and calculated the effective refractive index as a result of the multiple reflections.

In Fig. 10, we compare the optical constants obtained by the heating model simulation to the experimental data measured at $t_{pp}$ = 12.8 ps for the fluence of 3.4 mJ/cm$^2$ (red) and 0.4 mJ/cm$^2$ (blue). Here, we define the pump-induced phase shift of the reflected THz *E*-field as $\theta(\omega)$ = arg($r_{\mathrm{On}}(\omega)/r_{\mathrm{Off}}(\omega)$), where $r_{\mathrm{on}}(\omega)$ and $r_{\mathrm{off}}(\omega)$ are the complex reflectivity when the pump pulse is "On" and "Off", respectively. By using the literature values of the pump penetration depth [5] and specific heat [98], the heating model well reproduces the pivotal experimental observations without any adjustable parameters: the redshift of the JPR in the reflectivity, which is more clearly identified in the negative reflectivity change $\Delta R/R$ around the equilibrium JPR frequency of 1.2 THz, and the decrease in the phase $\theta(\omega)$ around the JPR frequency in equilibrium, both of which are attributed to the destruction of the superconductivity. The $\Delta R/R$ spectrum obtained by the experiment in Fig. 10(b) is slightly different from that obtained by the simulation in Fig. 10(e) above 1.4 THz, probably due to the QP excitation effect, which is not considered in the heating model simulation. We note that the heating model simulation is not applicable at earlier pump-probe delay times where the electron-lattice system does not reach the quasi-equilibrium state.

### APPENDIX D: DESTROYED LAYER MODEL

To elucidate the optical properties within ~ 6 ps after photoexcitation when the hating model is not applicable, we develop the following "destroyed layer model" as schematically illustrated in Fig. 11(a). Here, we assume that the superconductivity in the photoexcited surface region with a depth of $d_0$ is destroyed and the optical constants are described by that at 60 K. Firstly, by considering the multiple reflections of the THz probe pulse inside the photoexcited surface region, we calculate the effective reflectivity $R_{\mathrm{dest}}$. Figure 11(b) shows that the THz reflectivity change spectra $\Delta R/R$ at 20 K measured at $t_{pp}$ = 3.2 ps are reasonably fitted with $(R_{\mathrm{dest}} - R_{\mathrm{eq}})/R_{\mathrm{eq}}$, where $d_0$ is the only fitting parameter and the $R_{\mathrm{eq}}$ denotes the equilibrium reflectivity.

Secondly, if the superconductivity is destroyed within the thickness of $d_0$ at the sample surface, the attenuated pump intensity at the sample depth of $z = d_0$ should reach the SC threshold energy $I_{\mathrm{th}}$,

$$F_0 \exp\left(-\frac{d_0}{d_{\mathrm{pump}}}\right) = F_{\mathrm{th}}. \tag{D1}$$

Thus, the destroyed layer length $d_0$ can be written as

$$d_0(I_0) = \begin{cases} d_{\mathrm{pump}} \ln\left(\dfrac{F_0}{F_{\mathrm{th}}}\right) + d_{\mathrm{offset}} & (F_0 \geq F_{\mathrm{th}}) \\ d_{\mathrm{offset}} & (0 \leq F_0 < F_{\mathrm{th}}). \end{cases} \tag{D2}$$

Here, we add an offset $d_{\text{offset}}$ to prevent $d_0$ going to zero for $F_0 < F_{\text{th}}$. The destroyed layer length $d_0$ obtained by the fitting in Fig. 11(b) is plotted as a function of the pump fluence $F_0$ in Fig. 11(c). We fit the experimentally obtained $d_0$ using Eq. (D2) with the adjustable fitting parameters of $F_{\text{th}}$ and $d_{\text{offset}}$ while fixing the 800-nm pump penetration depth $d_{\text{pump}} = 0.22$ μm [5]. One can see that the obtained $d_0$ follows the fitting curve by Eq. (D2) and the SC threshold energy is estimated as $F_{\text{th}} = 70$ μJ/cm$^2$, whose energy scale is in good agreement with the results of the previous time-resolved pump-probe measurements [55-59] and ARPES measurements in cuprate superconductors [60-62]. This good agreement between the experimental results and the destroyed layer model reveals that, below $T_c$, the superconductivity is destroyed by the photoexcitation even before the system reaches the thermal quasi-equilibrium state.

**APPENDIX E: ANALYSIS OF THE SURFARCE OPTICAL RESPONSE AT 100 K**

In this Appendix, we present the details of the analysis to extract the transient optical response at 100 K. Figure 12(a) shows the raw data of the 800-nm pump-induced THz reflectivity change $\Delta R/R$ at 100 K measured at $t_{pp} = 0.8$ ps (red) and 9.6 ps (blue) with the pump fluence of 3.4 mJ/cm$^2$. The $\Delta R/R$ spectrum at $t_{pp} = 0.8$ ps exhibits a plasma edge-like behavior similar to the equilibrium JPR below $T_c$ in accordance with the previous studies [3-5,26]. We plot the reflectivity change calculated by the heating model simulation at 100 K in Fig. 12(a) by the blue dashed curve. Notably, the reflectivity change expected from the heating model simulation is much smaller than the experimental data, showing that the non-thermal effect plays a crucial role above $T_c$.

Figure 12(b) presents the spatial distribution of the pump fluence (magenta, left axis) and the sample temperature (light blue, right axis) obtained by the heating model simulation. The temperature rise at 100 K with the pump fluence of 3.4 mJ/cm$^2$ is limited within a small volume near the sample surface with its length scale in the depth direction determined by the pump penetration depth $d_{\text{pump}}$ because the specific heat is much larger than that below $T_c$. Therefore, even though the surface layer models extracting the optical constants in the photoexcited surface region might cause artifacts below $T_c$ as mentioned above, they serve as reasonable models above $T_c$ where the refractive index exhibits a monotonic temperature dependence, as shown in Figs. 9(b) and (c) in Appendix C. Similar models have been utilized to investigate the transient THz optical constants in photoexcited semiconductors [99]. In the following, we present the optical constants at the photoexcited surface region extracted by the so-called single layer model: we suppose that the refractive index depends on the sample depth $z$ as

$$n^{\text{SL}}(\omega, z) = \begin{cases} n_{\text{surf}}(\omega), & (0 \leq z \leq d_{\text{pump}}), \\ n_{\text{eq}}(\omega), & (z > d_{\text{pump}}) \end{cases}. \quad \text{(E1)}$$

Here, $n_{\text{surf}}(\omega)$ and $n_{\text{eq}}(\omega)$ are the complex refractive index at the photoexcited surface region and that in equilibrium. By considering the multiple reflections inside the photoexcited surface region, we compute the total optical response of the sample observed by the THz probe pulse $n_{\text{eff}}(\omega)$. Finally, we obtain the refractive index at the surface $n_{\text{surf}}(\omega)$ such that $n_{\text{eff}}(\omega)$ reproduces the experimentally observed complex refractive index.

## APPENDIX F: PUMP POLARIZATION DEPENDENCE IN THE C-AXIS TRANSIENT OPTICAL RESPONSES

We also performed the OPTP experiments by setting the 800-µm pump polarization along the *a*-axis while keeping the THz probe polarization along the *c*-axis. Figure 13 plots the 800-nm pump-induced transient reflectivity change $\Delta R/R$ along the *c*-axis of the YBCO at 20 K. $\Delta R/R$ can be fitted by the destroyed layer model described in Appendix D using the parameter of $d_0 = 0.43$ µm. This value of $d_0$ is longer than the 800-nm pump penetration depth of 0.11 µm along the *a*-axis, which is computed from the optical constants in the literature [100], as expected in the destroyed later model.

In Fig. 14, we present the transient increase in the *c*-axis optical conductivity at the photoexcited surface at 100 K when the pump polarization is parallel to the *c*-axis (red) and *a*-axis (green) using the single layer model in Eq. (E1). Remarkably, both the real and imaginary parts of the optical conductivity increase for the *a*-axis pump similar to the case of the *c*-axis pump. These results reveal that the NIR pump-induced transient increase in $\sigma_1(\omega)$ and $\sigma_2(\omega)$ does not depend on the pump polarization, contrary to the previous work with the MIR photoexcitation where the transient increase in $\sigma_2(\omega)$ was discerned only for the *c*-axis pump above $T_c$ [26].

## APPENDIX G: EVALUATION OF THE TEMPERATURE DEPENDENCE OF THE THz-THG INTENSITY

In this appendix, we describe the procedure to evaluate the temperature dependence of the THz THG intensity arising from the ac-driven Josephson current. First, we remove the leakage of the FH component in the following manner. Since the FH leakage intensity at TH frequency ($I_{\text{leak}}$) follows the squared incident FH *E*-field $E_{\text{in}}^2$ and the third-order nonlinear TH intensity ($I_{\text{THG}}$) follows $E_{\text{in}}^6$, as shown in Figs. 4(b) and (c), the measured TH intensity ($I_{\text{meas}}$) is expressed as a function of the incident FH *E*-field $E_{\text{in}}$ by

$$I_{\text{meas}}(E_{\text{in}}) = I_{\text{THG}}(E_{\text{in}}) + I_{\text{leak}}(E_{\text{in}}) = (\chi^{(3)} E_{\text{in}}^3)^2 + \alpha_{\text{leak}} E_{\text{in}}^2. \tag{G1}$$

Here, $\chi^{(3)}$ is the third-order nonlinear susceptibility, and $\alpha_{\text{leak}}$ is a coefficient for the leakage. The leakage component can be subtracted by measuring the TH intensity for two different incident FH $E$-fields of $E_{\text{in}}$ and $\beta E_{\text{in}}$, where the coefficient $\beta$ is controlled by wire grid polarizers in the path of the incident THz beam. In this case, the nonlinear TH intensity $I_{\text{THG}}(E_{\text{in}})$ can be written as

$$I_{\text{THG}}(E_{\text{in}}) = \frac{I_{\text{meas}}(E_{\text{in}}) - I_{\text{meas}}(\beta E_{\text{in}})/\beta^2}{1-\beta^4} \tag{G2}$$

During the measurements, $\beta$ is set to 0.75. Figure 15(a) shows the obtained TH intensity $I_{\text{THG}}$ as a function of temperature (magenta curve). We note that $I_{\text{THG}}$ is the TH intensity integrated from 1.2 to 1.8 THz after subtracting the leakage. The TH intensity $I_{\text{THG}}$ displays a sharp increase below $T_c$ and a maximum at 52 K. This peak at 52 K is attributed to the screening effect of the THz $E$-field which exhibits non-monotonic temperature dependence, as we discuss below.

The screening effect of the incident FH $E$-field and outgoing TH $E$-field from the sample is evaluated using the complex refractive index $n(\omega)$ obtained by the THz-TDS measurements. The situation is depicted in Fig. 15(b). The screened FH $E$-field inside the sample $E_{\text{scr}}(\omega_0)$ is related to the incident FH $E$-field $E_{\text{in}}(\omega_0)$ as

$$E_{\text{scr}}(\omega_0) = \frac{2}{1+n(\omega_0)} E_{\text{in}}(\omega_0) = t_{12}(\omega_0) E_{\text{in}}(\omega_0). \tag{G3}$$

Here, $t_{12}(\omega_0)$ is the Fresnel transmission coefficient for the FH angular frequency $\omega_0$ from the air to the sample. The FH $E$-field inside the sample $E_{\text{scr}}(\omega_0)$ generates the TH $E$-field $E_{\text{sample}}(3\omega_0) = \chi^{(3)} E_{\text{scr}}(\omega_0)^3$. Thus, the observed TH $E$-field $E_{\text{THG}}(3\omega_0)$ is written by

$$E_{\text{THG}}(3\omega_0) = \frac{2n(3\omega_0)}{1+n(3\omega_0)} E_{\text{sample}}(3\omega_0) = t_{21}(3\omega_0)\chi^{(3)} E_{\text{scr}}(\omega_0)^3, \tag{G4}$$

where $t_{21}(3\omega_0)$ is the Fresnel transmission coefficient for the angular frequency $3\omega_0$ from the sample to the air. Using Eqs. (G3) and (G4), the third-order nonlinear susceptibility $\chi^{(3)}$ is expressed as

$$\chi^{(3)} = \frac{E_{\text{THG}}(3\omega_0)}{t_{12}(\omega_0)^3 E_{\text{in}}(\omega_0)^3 t_{21}(3\omega_0)} = \sqrt{\frac{1}{B_{\text{THG}}(\omega_0)} \frac{I_{\text{THG}}(3\omega_0)}{I_{\text{in}}(\omega_0)^3}}, \tag{G5}$$
$$B_{\text{THG}}(\omega_0) = t_{12}(\omega_0)^3 t_{21}(3\omega_0).$$

Here, $I_{\text{in}}(\omega_0)$ and $I_{\text{THG}}(3\omega_0)$ are the intensity of the incident FH $E$-field and observed TH $E$-field, respectively. The coefficient $B_{\text{THG}}$ at 0.5 THz is computed using the refractive index $n(\omega)$ and plotted as the light blue dashed curve in Fig. 15(a).

Finally, we consider that the Josephson current induced by the FH incident field is proportional to the area perpendicular to the $c$-axis where the FH field penetrates inside the sample, because the nonlinear polarization is induced within the volume of the penetration depth from the surface by the THz pulse in reflection geometry [69,70]. If the FH beam area on the sample is fixed for all the temperatures, this volume is proportional to the FH penetration depth, which can be calculated as $d_{THz}(\omega_0) = c/[2n_2(\omega_0)\omega_0]$. It is convenient to define the third-order nonlinear susceptibility per depth $\chi_{dep}^{(3)} = \chi^{(3)}/d_{THz}$ when we consider the expected TH intensity at 100 K after photoexcitation. We present the obtained $\chi_{dep}^{(3)}$ as a function of temperature in Fig. 4(d).

## APPENDIX H: SIMULATION OF THE THz-THG INTENSITY AFTER PHOTOEXCITATION

We simulate the time evolution of the THz-THG intensity driven by the Josephson current under the 800-nm photoexcitation by applying the procedure reported in Ref. [46] to describe the transient THG response associated with the SC order parameter. When a THz pulse polarized along the c-axis of YBCO ($E(t) = E_0 \sin(\omega_0 t)$) is irradiated, the interlayer phase difference of the SC order parameter $\theta(t)$ advances in time according to the Josephson relation as

$$\frac{\partial \theta(t)}{\partial t} = \frac{2ed}{\hbar} E(t). \quad (H1)$$

Here, $2e$ is the Cooper pair charge, $d$ is the interlayer spacing, and $\hbar$ is the Planck's constant $h$ divided by $2\pi$. Following the time evolution of $(t) = \theta_0 \cos(\omega_0 t)$ ($\theta_0 = 2edE_0/\hbar$), the Josephson current $I(t)$ is induced and given by

$$I(t) = I_c \sin \theta(t) \approx I_c[\theta_0 \cos(\omega_0 t) - \frac{\theta_0^3}{6} \cos^3(\omega_0 t)], \quad (H2)$$

where $I_c$ is the critical current and proportional to the superfluid density $n_s$ [101]. For an intense THz $E$-field, the second term in the right side of Eq. (H2) gives rise to the THG.

Now we consider the time evolution of the THz-THG intensity by assuming that the 800-nm photoexcitation induces the superconductivity. In the OP-THG experiment, the THz $E$-field was measured by sweeping the THz pulse arrival time at the sample $t_{probe}$ while fixing the timing between the pump and the sampling pulse for the EO-sampling $t_{pp} = t_{samp} - t_{pump}$, where $t_{samp}$ is the timing of the sampling pulse for the EO-sampling and $t_{pump}$ is the time when the 800-nm pump pulse arrives at the sample. This configuration ensures that the THz $E$-field waveform is composed of the data points measured at the same delay time $t_{pp}$ [4,102-107]. Here, we assume that the superfluid density $n_s$ is induced after 800-nm photoexcitation and advances in time $t_{pp}$ as shown in the inset of Fig. 6(d). In the simulation, we employ the THz $E$-field used in the experiment

$E(t_\text{samp} - t_\text{probe})$. Since the critical current $I_c$ is proportional to $n_s(t_\text{samp} - t_\text{pump})$, Eq. (H2) can be rewritten as

$$I^\text{wp}(t_\text{samp} - t_\text{pump}, t_\text{samp} - t_\text{probe}) \propto n_s(t_\text{samp} - t_\text{pump}) \sin\theta(t_\text{samp} - t_\text{probe}). \qquad (\text{H3})$$

Here, $t_\text{probe}$ is the time when the THz pulse arrives at the sample. The emitted $E$-field from this Josephson current is expressed as

$$E^\text{wp}(t_\text{samp} - t_\text{pump}, t_\text{samp} - t_\text{probe}) = -\frac{\partial}{\partial t_\text{samp}} I^\text{wp}(t_\text{samp} - t_\text{pump}, t_\text{samp} - t_\text{probe}). \qquad (\text{H4})$$

By sweeping $t_\text{probe}$ while keeping $t_{pp} = t_\text{samp} - t_\text{pump}$, we can calculate $E^\text{wp}$ for each $t_{pp}$. The simulated TH intensity of $E^\text{wp}$ is shown as a function of $t_{pp}$ in Fig. 6(d).

**Figures**

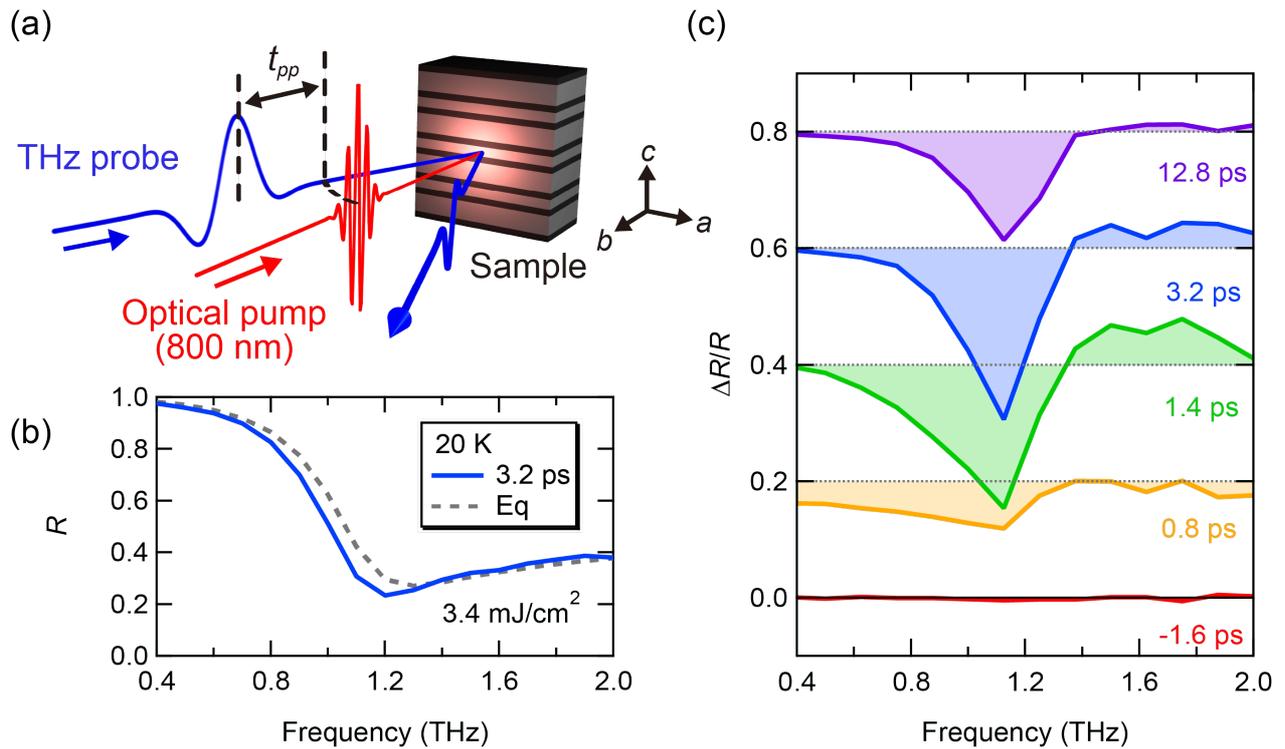

Figure 1 (a) Schematic illustration of the optical pump-THz probe (OPTP) spectroscopy. (b) The 800-nm pump-induced transient reflectivity along the $c$-axis of the YBCO at 20 K with the pump fluence of 3.4 mJ/cm$^2$ (solid blue curve). The pump polarization is set parallel to the $c$-axis. The dashed gray curve represents the equilibrium reflectivity spectrum at 20 K without the pump pulse. (c) The 800-nm pump-induced transient reflectivity change at 20 K with the pump fluence of 3.4 mJ/cm$^2$ at selected pump-probe delay times $t_{pp}$. The delay time $t_{pp}$ for each data is shown in the right.

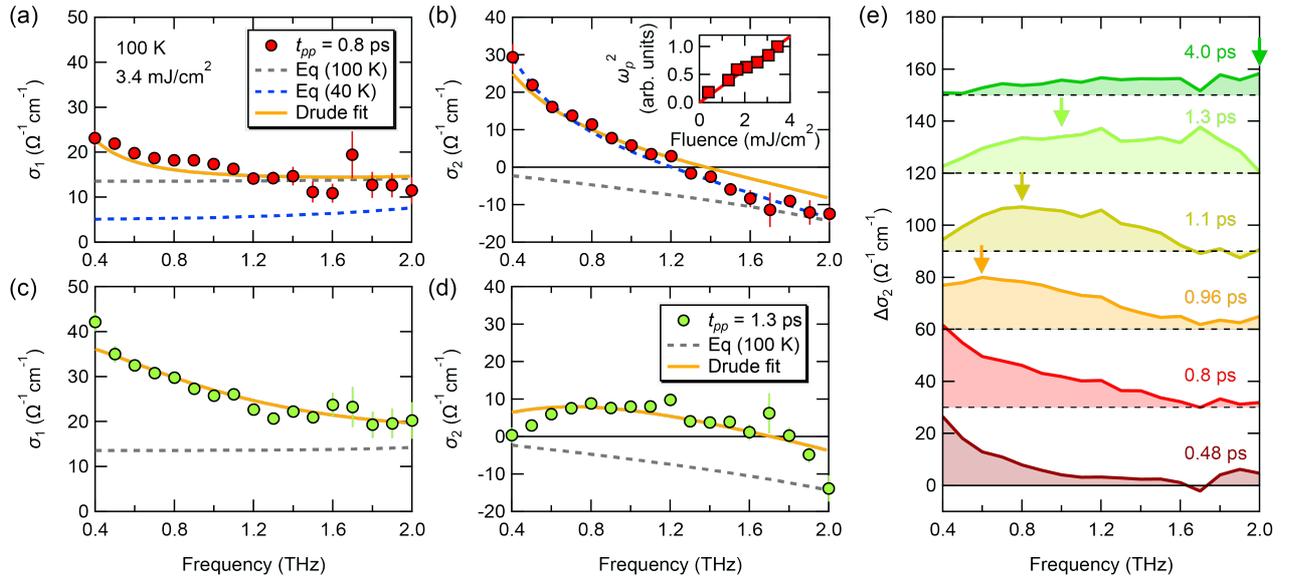

Figure 2 (a), (b) The real and imaginary parts of the transient optical conductivity along the c-axis of the YBCO at 100 K with the 800-nm pump fluence of 3.4 mJ/cm$^2$ measured at $t_{pp}$ = 0.8 ps (red circles). Here, the pump polarization is along the c-axis. The dashed curves represent the spectra in equilibrium without the pump. The solid orange curves are the fits to the data using the Drude model. The inset in (b) presents the fluence dependence of the squared plasma frequency $\omega_p^2$ evaluated by the Drude fit at $t_{pp}$ = 0.8 ps. The solid line is the linear fit to the data. (c), (d) The real and imaginary parts of the transient optical conductivity along the c-axis at 100 K with the pump fluence of 3.4 mJ/cm$^2$ measured at $t_{pp}$ = 1.3 ps (green circles). The solid orange curves are the fits to the data using the Drude model (e) The time evolution of the pump-induced change in the imaginary part of the optical conductivity at 100 K with the 800-nm pump fluence of 3.4 mJ/cm$^2$. The delay time $t_{pp}$ for each data is shown in the right. The vertical arrows denote the peak positions in $\Delta\sigma_2(\omega)$.

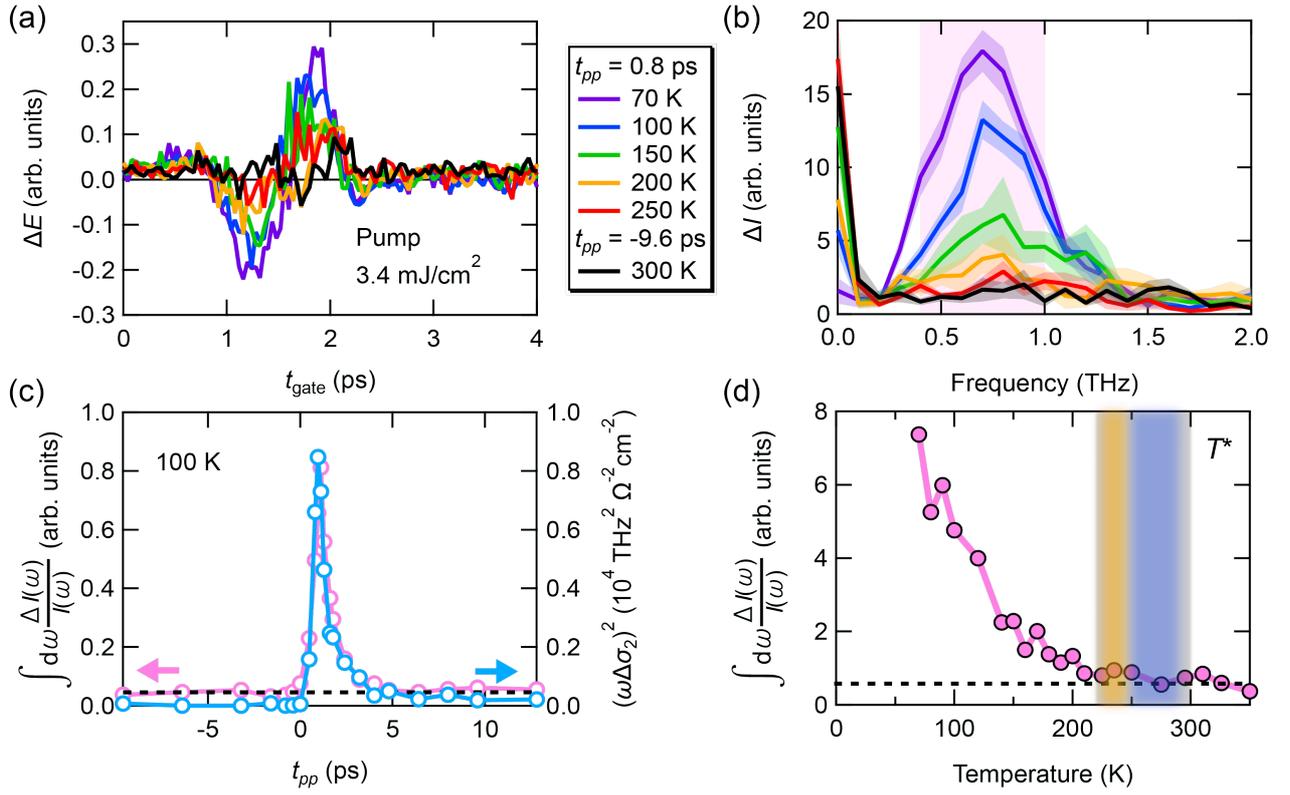

Figure 3 (a) The 800-nm pump-induced change in the reflected THz *E*-field waveform along the *c*-axis of the YBCO (denoted by Δ*E*(*t*$_{gate}$)) with the pump fluence of 3.4 mJ/cm$^2$ measured at the indicated temperatures above *T*$_c$. The pump polarization is parallel to the *c*-axis. (b) The FFT intensity of Δ*E*(*t*$_{gate}$) (denoted by Δ*I*(ω)) corresponding to (a). The standard errors of the measurements are displayed as colored bands. (c) The integrated intensity of Δ*I*(ω) normalized by the reflected THz power as a function of the pump-probe delay time *t*$_{pp}$ measured at 100 K (magenta, left axis). The integral region is shown by the shaded area (magenta) in (b). The horizontal black dashed line denotes the noise floor estimated from the integrated FFT intensity at the negative delay time. The squared pump-induced change in the imaginary part of the optical conductivity multiplied by the angular frequency, (ωΔσ$_2$)$^2$, at ω/2π = 0.4 THz is also shown in the right axis (light blue). (d) The integrated FFT intensity of Δ*I*(ω) normalized by the reflected THz power without photoexcitation as a function of temperature. The horizontal dashed black line indicates the noise floor estimated from the integrated FFT intensity at 300 K measured at *t*$_{pp}$ = −9.6 ps. The pseudogap opening temperature *T*$^*$ of similarly doped YBCO are shown by the vertical shaded lines reported in neutron scattering (235±10 K, orange) [64], in Nernst measurement (260±40 K, blue) [65], and in optical spectroscopy measurement (270±20 K, gray) [66].

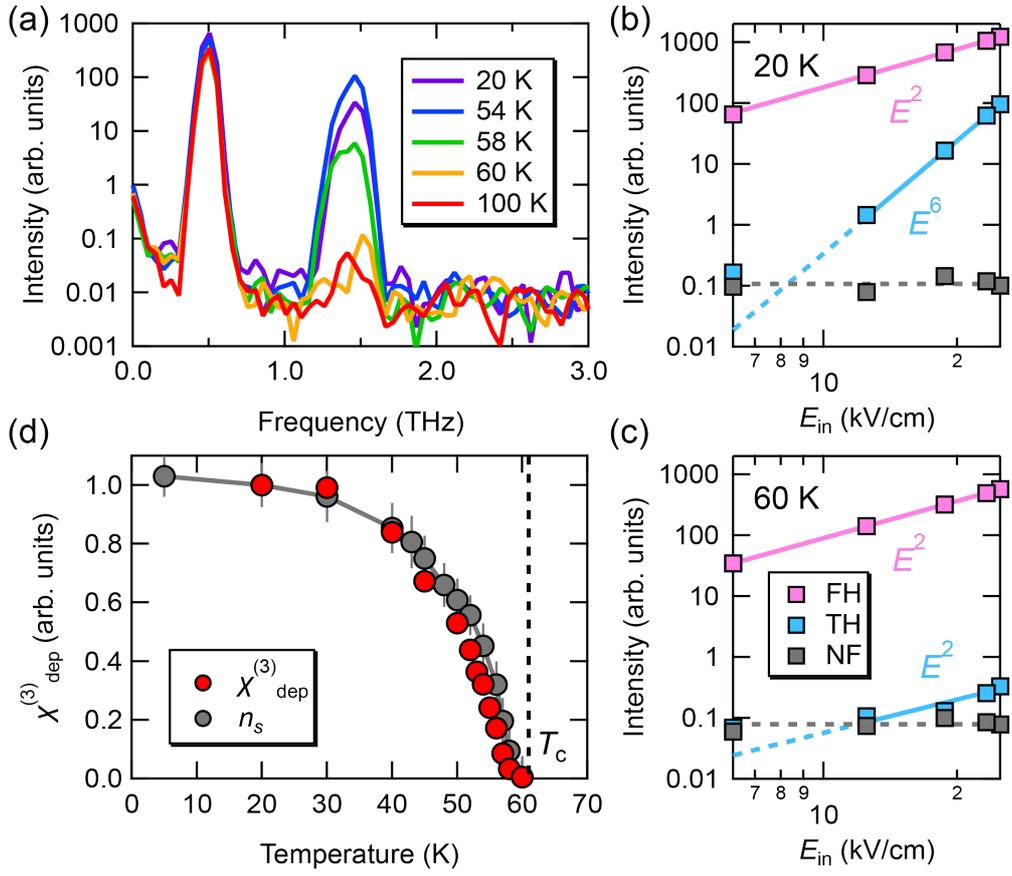

Figure 4 (a) The FFT power spectra of the reflected narrowband THz $E$-field from the YBCO at selected temperatures using the THz peak $E$-field of 25 kV/cm. (b) The frequency-integrated FH (magenta squares) and TH (light blue squares) intensity as a function of the incident THz peak $E$-field ($E_{in}$) at 20 K. The FH intensity is integrated from 0.3 to 0.7 THz, and the TH intensity is integrated from 1.2 to 1.8 THz. The solid lines are the guides to the eye with a slope of 2 (magenta line) and 6 (light blue line). The gray squares show the noise floor, which are estimated by integrating the FFT power spectra as in (a) from 2.5 to 3.1 THz. The gray dashed line denotes the fitting curve to the noise-floor intensities with a constant. (c) The same plot as (b), but at 60 K. The solid lines are the guides to the eye with a slope of 2. In the legend, NF denotes the noise floor. (d) Temperature dependence of the third-order nonlinear susceptibility per depth $\chi^{(3)}_{dep}$ (red) and the superfluid density $n_s$ (gray). The error bars for $n_s$ are the fitting errors in the two-fluid model. The black vertical dashed line denotes $T_c$.

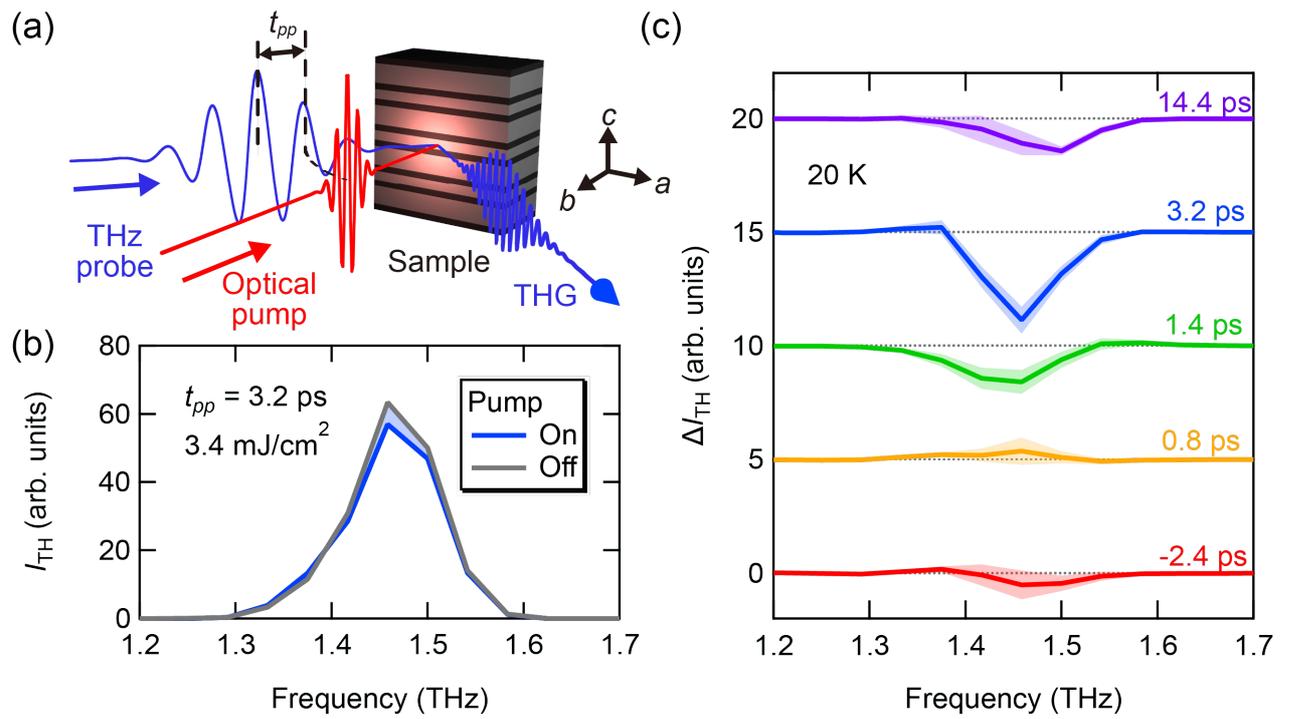

Figure 5 (a) Schematic of the optical pump-THG probe (OP-THG) spectroscopy in reflection geometry. Here, both the 800-nm pump and 0.5-THz probe polarizations are parallel to the $c$-axis of the YBCO. (b) The FFT power spectra of the reflected narrowband THz $E$-field at 20 K measured at $t_{pp}$ = 3.2 ps when the 800-nm pump pulse is "On" (blue curve) and "Off" (gray curve). (c) The 800-nm pump-induced change in the TH intensity at 20 K with the pump fluence of 3.4 mJ/cm$^2$ at selected pump-probe delay times $t_{pp}$. The delay time $t_{pp}$ for each data is shown in the right.

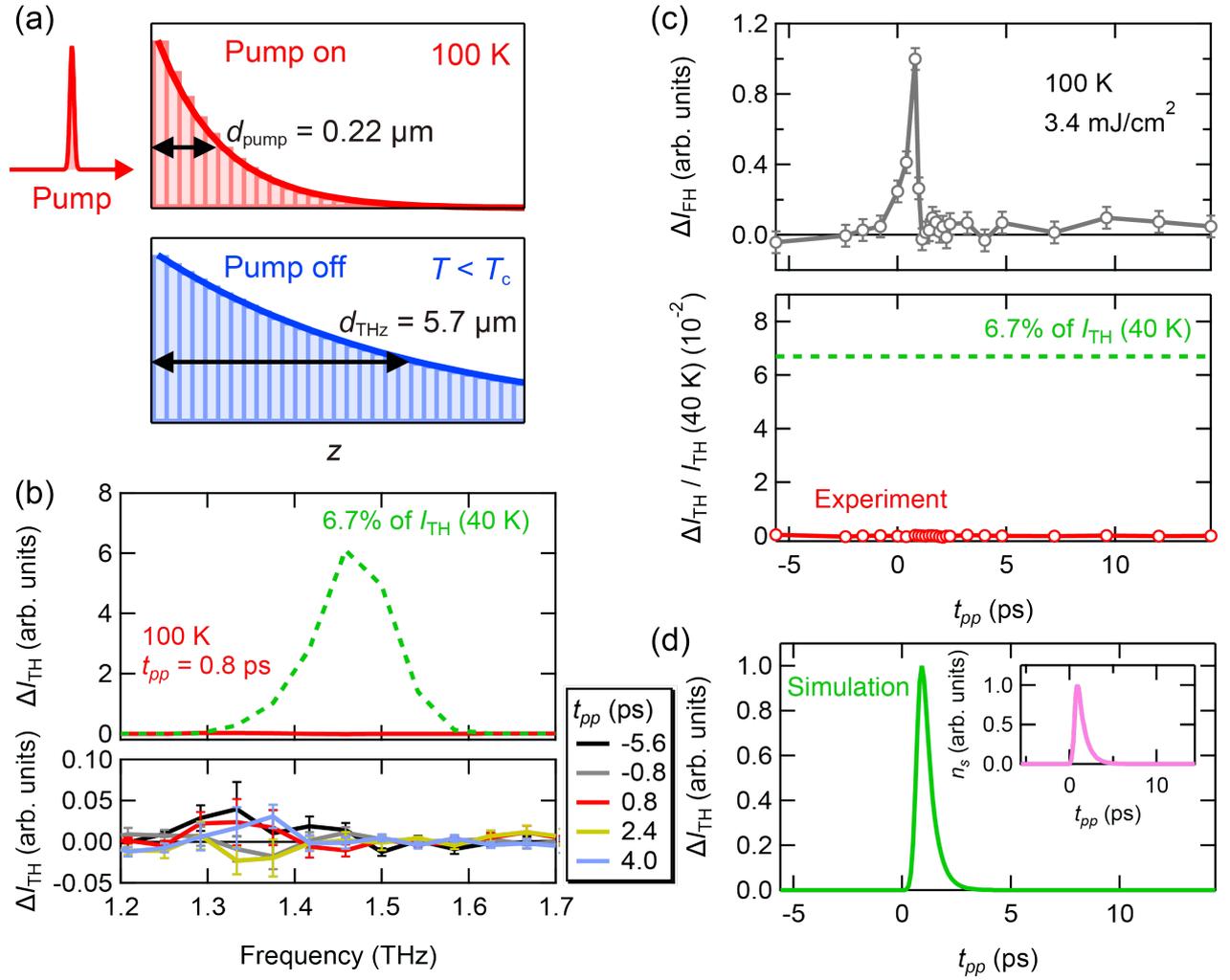

Figure 6 (a) Schematic illustration of the penetration depth of the optical pump above $T_c$ at 100 K compared to that for the THz probe in the SC phase below $T_c$. (b) The upper panel shows the pump-induced change in the FFT power spectrum around TH frequency for the YBCO at 100 K (red curve). The green dashed curve is 6.7% of the FFT power spectrum at 40 K without the photoexcitation. Both the 800-nm pump and 0.5-THz probe polarizations are along the $c$-axis of the YBCO. The lower panel shows the expanded plot of the pump-induced change in FFT power spectrum at 100 K measured at the selected delay times. The error bars denote the standard errors of the measurements. (c) The pump-induced change in the FH intensity ($\Delta I_{FH}$, upper panel) and TH intensity ($\Delta I_{TH}$, lower panel) at 100 K as a function of the pump-probe delay time $t_{pp}$. The FH intensity changes are integrated from 0.3 to 0.7 THz, and the TH intensity change is integrated from 1.2 to 1.8 THz. In the lower panel, $\Delta I_{TH}$ at 100 K is normalized by the integrated TH intensity without photoexcitation at 40 K from 1.2 to 1.8 THz, and 6.7% of the integrated TH intensity at 40 K is shown by the green dashed line. (d) Simulated time evolution of the TH intensity expected for the light-induced SC state. The inset plots the time evolution of the superfluid density $n_s$ after photoexcitation normalized by its maximum value, which is used in the simulation (Appendix H).

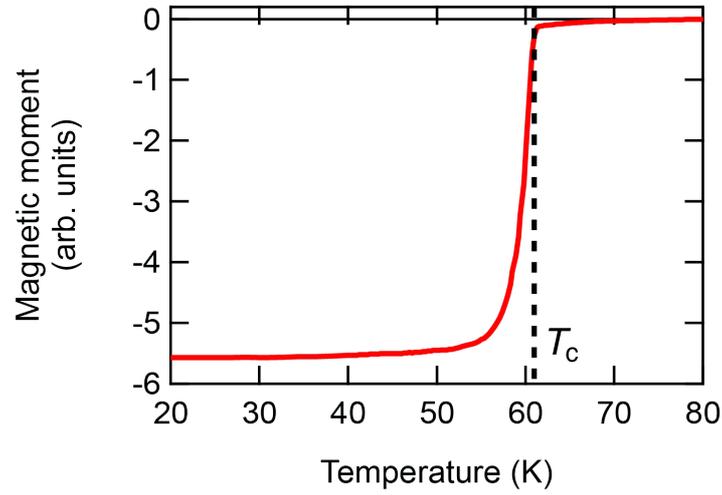

Figure 7 The magnetic moment of the underdoped YBCO single crystal. The black vertical dashed lines denote the determined $T_c$.

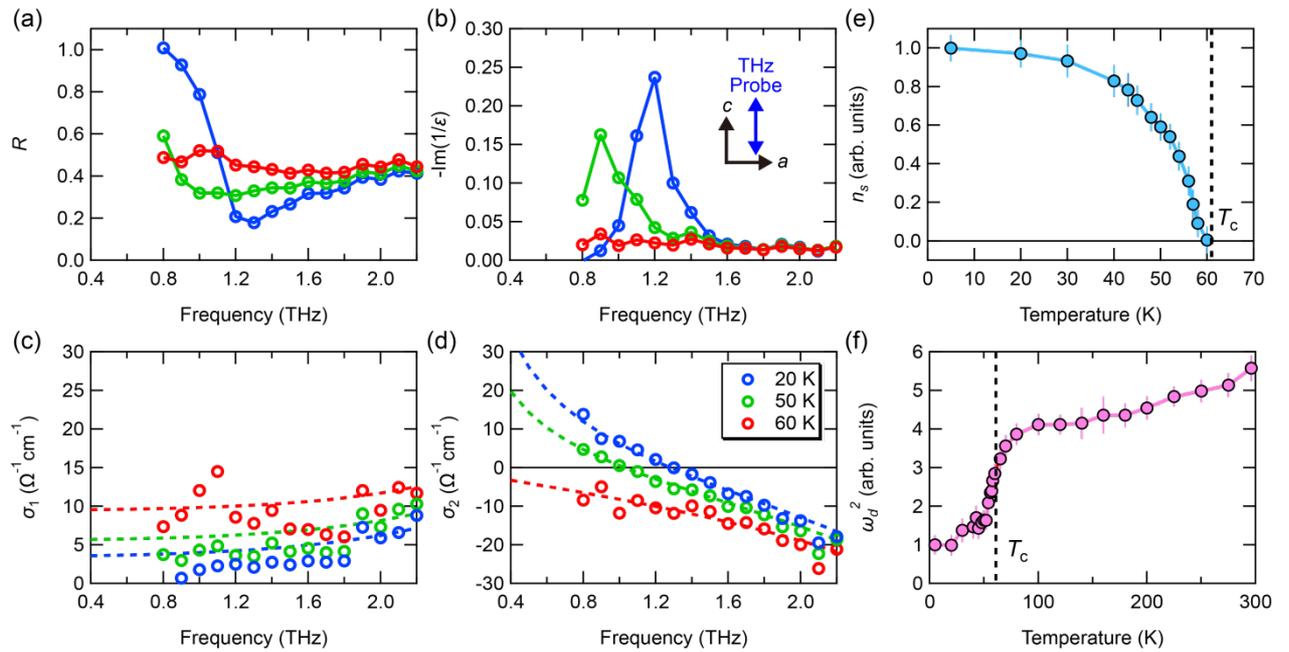

Figure 8 The *c*-axis equilibrium optical constants of the YBCO obtained by THz time-domain spectroscopy (THz-TDS): (a) the reflectivity, (b) loss function, (c) real and (d) imaginary part of the optical conductivity. Fitting curves of the real and imaginary parts of the optical conductivity using the two-fluid model are shown by the dashed curves in (c) and (d), respectively. (e), (f) The fitting parameters of $n_s$ and $\omega_d^2$ as a function of temperature. The error bars are the fitting errors. The black vertical dashed lines denote $T_c$.

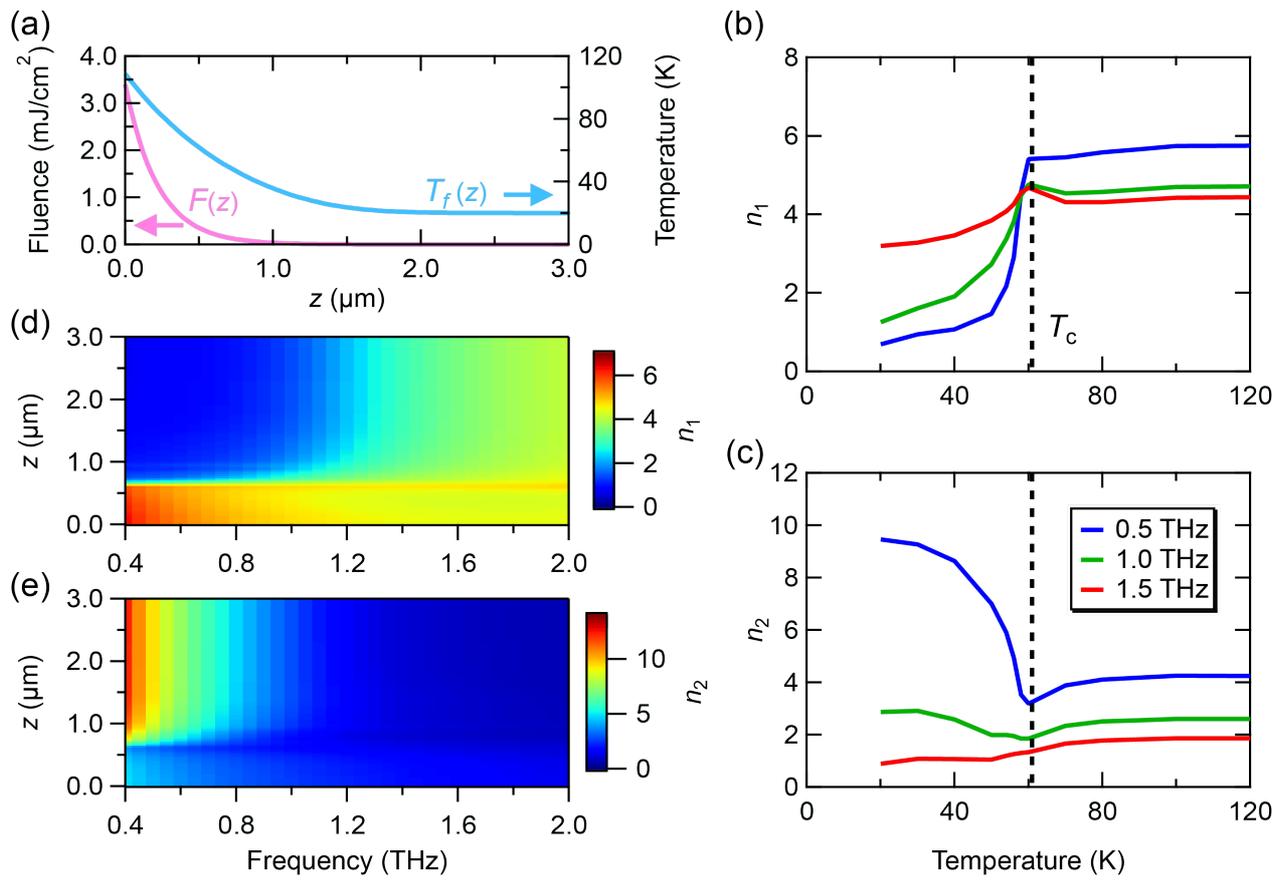

Figure 9 (a) The spatial distribution of the pump fluence $F(z)$ (magenta, left axis) and the sample temperature $T_f(z)$ (light blue, right axis) computed by the heating model simulation for the initial sample temperature of $T_i = 100$ K. (b), (c) Temperature dependence of the real and imaginary parts of the *c*-axis refractive index of the YBCO in equilibrium. The black vertical dashed lines denote $T_c$. (d), (e) Spatial distribution of the real and imaginary parts of the complex refractive index $n(\omega, z)$ constructed in the heating model simulation.

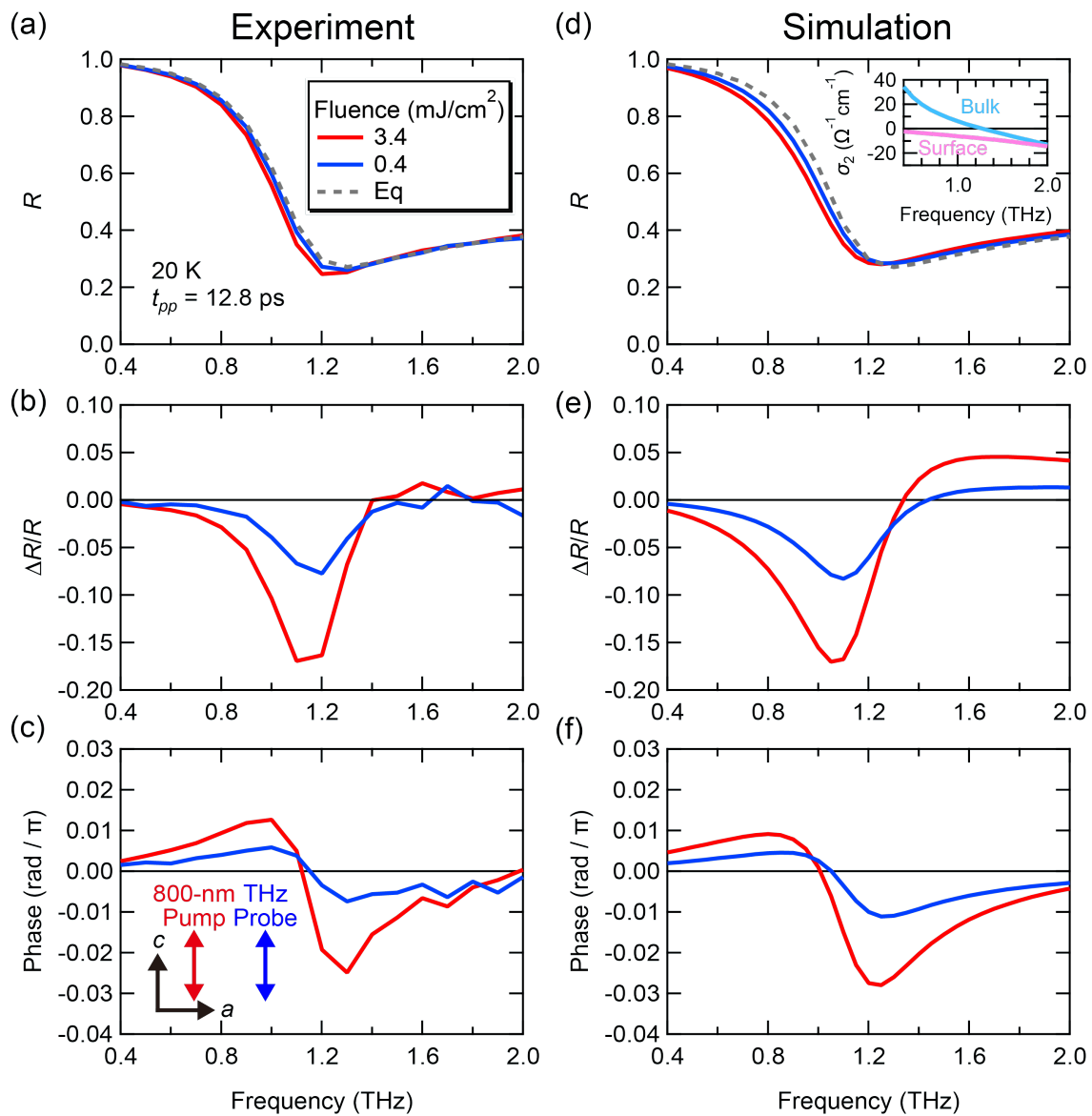

Figure 10 Comparison of the optical properties in the photoexcited nonequilibrium state: (a-c) the experimental results and (d-f) the calculated spectrum obtained by the heating model simulation. (a) The pump-induced transient reflectivity, (b) reflectivity change Δ$R$/$R$, and (c) phase shift $\theta$ at 20 K measured at $t_{pp}$ = 12.8 ps. Here, both the pump and probe polarizations are parallel to the $c$-axis of the YBCO. (d), (e) and (f) are the corresponding results computed by the heating model simulation. The inset in (d) shows the imaginary part of the optical conductivity at the surface (magenta) and the bulk (light blue) used in the heating model simulation for the pump fluence of 3.4 mJ/cm$^2$.

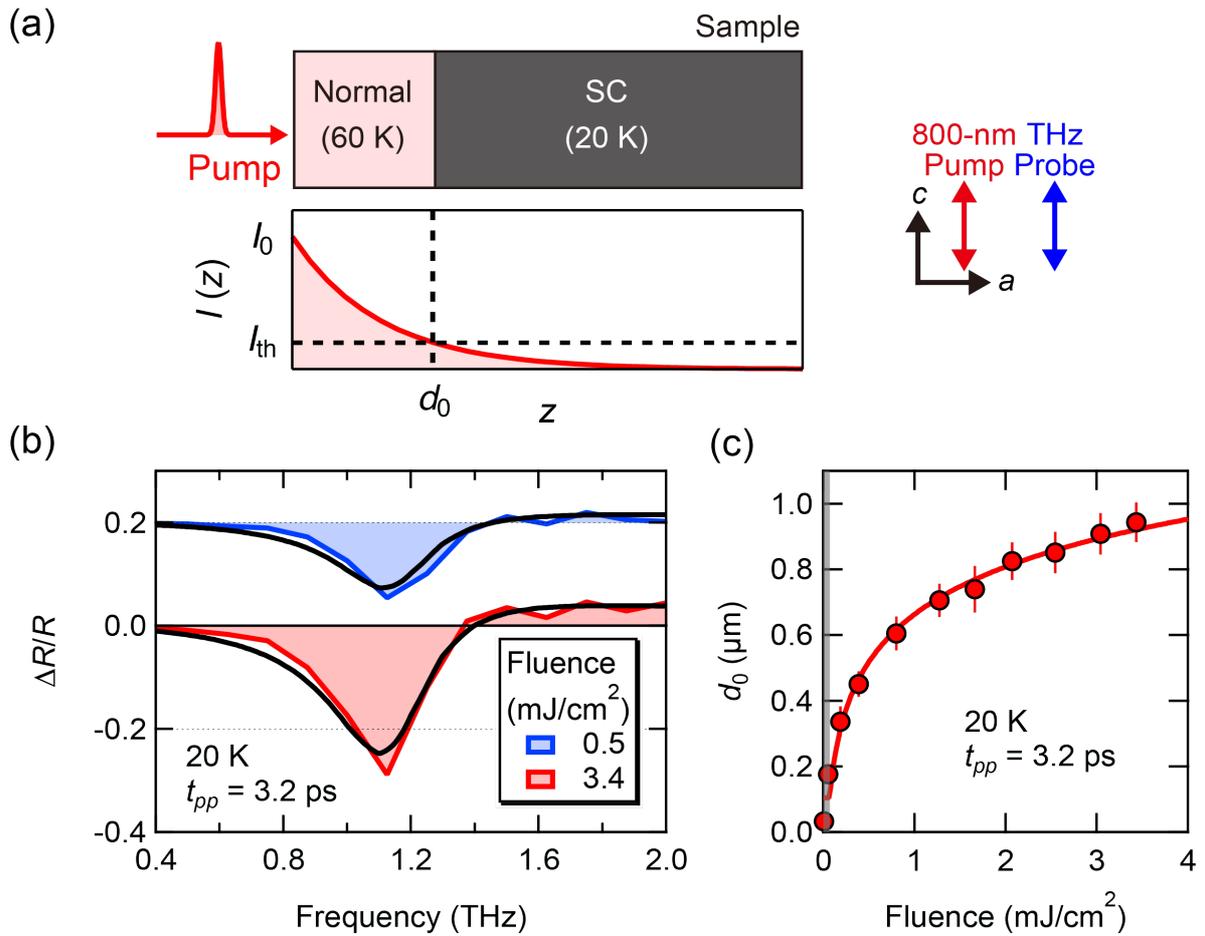

Figure 11 (a) Schematic illustration of the destroyed layer model presented in Appendix D. (b) The 800-nm pump-induced THz reflectivity change $\Delta R/R$ along the $c$-axis of the YBCO at 20 K. The black curves are the fits to $\Delta R/R$ using the destroyed layer model. (c) The destroyed layer length $d_0$ obtained by the destroyed model as a function of the 800-nm pump fluence at 20 K (red circles). The solid red curve is the fit to the data with Eq. (D2). The fluence lower than the determined threshold energy $F_{th}$ is shaded by gray.

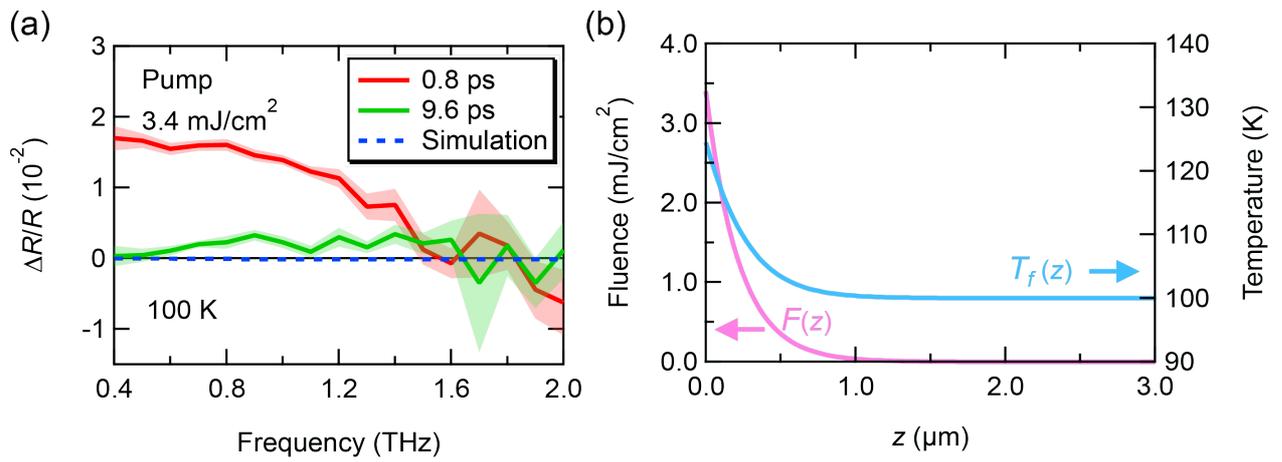

Figure 12 (a) The 800-nm pump-induced reflectivity change along the *c*-axis of the YBCO at 100 K with the pump fluence of 3.4 mJ/cm². The pump polarization is set parallel to the *c*-axis. The error bars are displayed as colored bands. The blue dashed curve shows the result of the heating model simulation at 100 K. (b) The spatial distribution of the pump fluence $F(z)$ (magenta, left axis) and the sample temperature $T_f(z)$ (light blue, right axis) calculated by the heating model simulation for the initial sample temperature of $T_i = 100$ K.

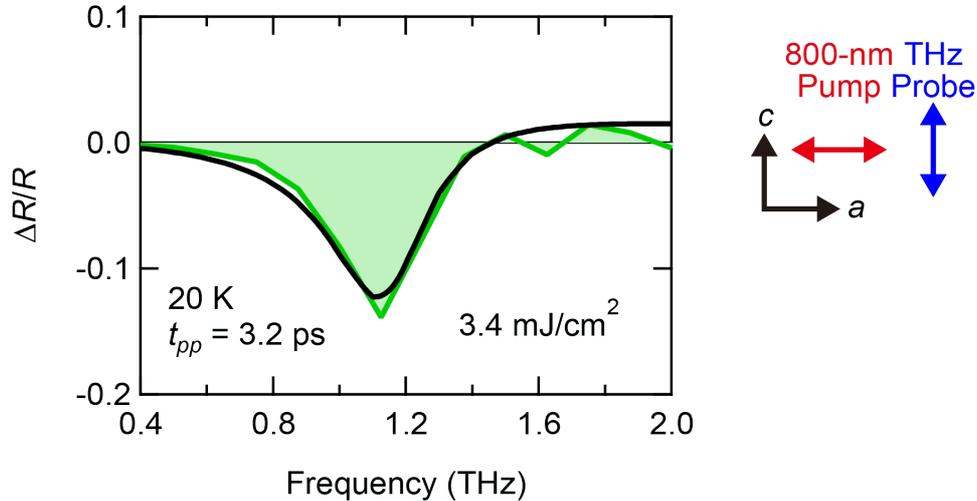

Figure 13 The 800-nm pump-induced THz reflectivity change $\Delta R/R$ along the *c*-axis of the YBCO at 20 K when the pump polarization is along the *a*-axis. The black curve is the fitting to $\Delta R/R$ using the destroyed layer model.

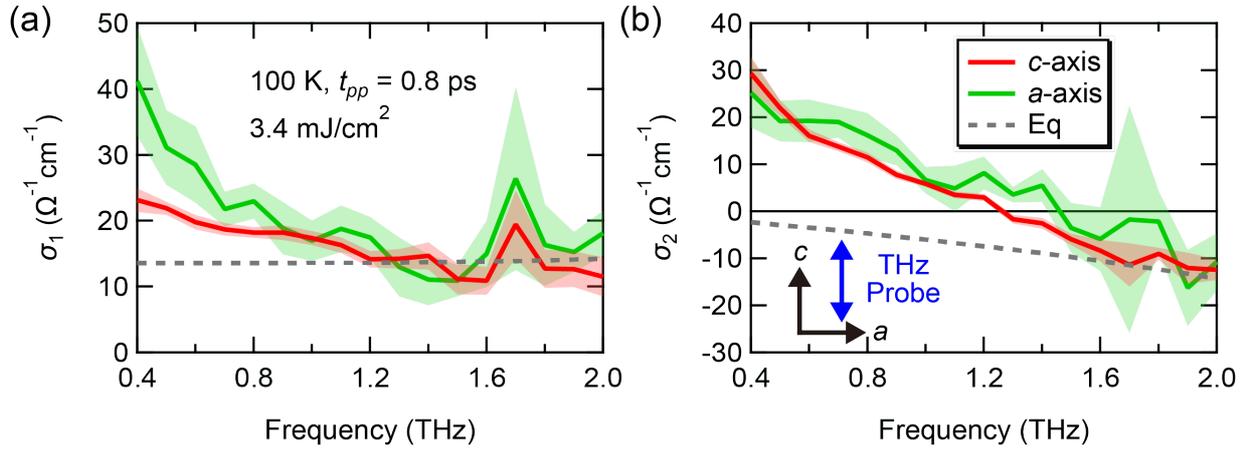

Figure 14 (a), (b) The real and imaginary parts of the transient optical conductivity along the $c$-axis of the YBCO at 100 K with the pump fluence of 3.4 mJ/cm$^2$ measured at $t_{pp}$ = 0.8 ps when the 800-nm pump polarization is parallel to the $c$-axis (red) and $a$-axis (green). The standard errors of the measurements are displayed as colored bands.

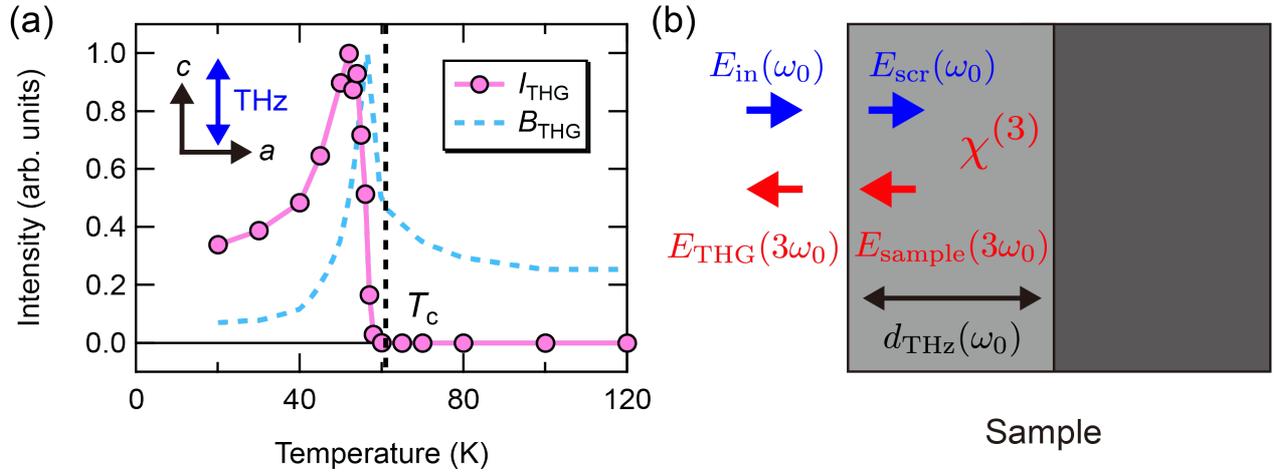

Figure 15 (a) Temperature dependence of the TH intensity from the YBCO after subtracting the leakage ($I_{THG}$, magenta curve). The light blue dashed curve is the screening coefficient $B_{THG}$ defined by Eq. (G5) at 0.5 THz as a function of temperature. The black vertical dashed line denotes $T_c$. (b) Schematic illustration of the THG in reflection geometry to consider the screening effect of the THz $E$-field inside the sample.